%% file: main.tex
\documentclass[conference]{IEEEtran}
\IEEEoverridecommandlockouts

\usepackage{url}
\usepackage{multirow}
\usepackage{subfigure}
\usepackage{epsfig}
\usepackage{graphicx}
\usepackage{amsmath}
\usepackage{pifont}
\usepackage{color}
\usepackage{setspace}
\usepackage{hhline}   
\usepackage{algorithm}
\usepackage{algpseudocode}
\usepackage{flushend}
\usepackage{url}
\usepackage[normalem]{ulem}
\usepackage{graphics}
\usepackage{tikz}
\usepackage{array}
\usepackage{animate}
\usepackage{titlesec}
\usepackage{balance}
\usepackage{enumitem}
\usepackage{booktabs}
\usepackage{amssymb}
\usepackage{amsmath}

\newcommand{\ignore}[1]{}

\usepackage{cite}
\usepackage{graphicx}
\usepackage{textcomp}
\usepackage{xcolor}

\begin{document}

\title{ML-based AIG Timing Prediction\\
to Enhance Logic Optimization}

\author{\IEEEauthorblockN{Wenjing Jiang}
\IEEEauthorblockA{
\textit{University of Minnesota}\\
Minneapolis, MN, USA}
\and
\IEEEauthorblockN{Jin Yan}
\IEEEauthorblockA{
\textit{Google DeepMind}\\
Mountain View, CA, USA}
\and     
\IEEEauthorblockN{Sachin S. Sapatnekar}
\IEEEauthorblockA{
\textit{University of Minnesota}\\
Minneapolis, MN, USA}
}

\maketitle

\begin{abstract}
As circuit designs become more intricate, obtaining accurate performance estimation in early stages, for effective design space exploration, becomes more time-consuming. Traditional logic optimization approaches often rely on proxy metrics to approximate post-mapping performance and area. However, these proxies do not always correlate well with actual post-mapping delay and area, resulting in suboptimal designs. To address this issue, we explore a ground-truth-based optimization flow that directly incorporates the exact post-mapping delay and area during optimization. While this approach improves design quality, it also significantly increases computational costs, particularly for large-scale designs. To overcome the runtime challenge, we apply machine learning models to predict post-mapping delay and area using the features extracted from AIGs. Our experimental results show that the model has high prediction accuracy with good generalization to unseen designs. Furthermore, the ML-enhanced logic optimization flow significantly reduces runtime while maintaining comparable performance and area outcomes.

\end{abstract}


\input{sec/1-introduction}
\input{sec/2-preliminary}

\input{sec/3-methodology}

\input{sec/4-experiment}

\input{sec/5-conclusion}

\vspace{4mm}
\section*{Acknowledgment}
This work is supported by Google DeepMind. The authors would like to thank Xiaoqing Xu, Dino Ruić, Raj B. Apte, Vinod Nair, Georges Rotival, Timur Sitdikov from Google DeepMind for all the helpful discussion and feedback.


\bibliographystyle{misc/ieeetr2}
\bibliography{references}

\end{document}

%% file: sec/1-introduction.tex
\section{Introduction}

\noindent
As technology scales down to smaller nodes, logic optimization has become more critical than ever before.  Suboptimal logic structures can lead to increased delay, area, and excessive power consumption. Several techniques for logic structure optimization have been extensively studied in previous works. Heuristic methods~\cite{Brayton84} may converge quickly, but could get stuck in local optima; SAT-based methods~\cite{Mishchenko05}~\cite{Sorensson05} perform an exhaustive search to find optimal solutions but are impractical for optimizing large designs due their high computational cost; simulated annealing (SA)~\cite{Kirkpatrick83,googleSA23} is effective in exploring the global solution space but may be computationally expensive;
genetic algorithms~\cite{Ohmori97}, based on evolutionary approaches, can be effective for complex problems,
but require more parameters and tuning effort than SA. 

In recent years, machine learning (ML) techniques have been applied in logic optimization with promising results. In~\cite{deAbreu21}, a decision tree-based approach has been applied to minimize the depth and the number of nodes in an and-inverter graph (AIG) logic representation. Neural networks have been employed in~\cite{Neto19} to choose appropriate logic optimizers for different parts of a circuit for modern designs that consist of heterogeneous blocks. 
Other works include \cite{Yu20}, which uses long short-term memory (LSTM) networks to predict design metrics for a given synthesis flow applied on a design, and \cite{Zheng24}, which focuses on predicting timing for a given optimization sequence, but under a limited range of logic transformations. A self-evolving system for synthesis parameter tuning~\cite{Ziegler16} is developed to automate solution space exploration. Logic optimization has been formulated as a reinforcement learning (RL) problem, as in~\cite{Haaswijk18}   (for majority-inverter graphs (MIGs)) and~\cite{Hosny20} (for AIGs), where RL has been explored for learning a better sequence of transformations to improve the quality of the design. In \cite{Haaswijk18}, only the depth of MIG is optimized in the proposed flow, while in~\cite{Hosny20}, the search space only consists of seven primitive transformations and the reward function only considers the direction of the delay and area change instead of quantifying how much change each action brings.

A major limitation of these prior approaches is that they utilize proxy metrics, e.g., taking the delay to be proportional to the logical depth  of the circuit,
and incorporate them into objective functions for optimization. However, proxy delay metrics may not be well correlated with the post-synthesis delay after technology mapping, and can be ineffective in optimizing the design.
To overcome these limitations of prior methods, this work studies how logic optimization can be improved by incorporating 
alternatives to approximate proxy metrics, to compute the delay in the cost function of a conventional optimization paradigm. These performance metrics must be computed efficiently to maintain the ability of logic synthesis to explore the large solution space efficiently, so that synthesis is scalable to large designs. 

In principle, it is possible to integrate the post-mapping delay in the cost function by performing the full technology mapping step. Our study finds that this can lead to a design of better quality than the optimization based on proxy metrics, but due to the high cost of the mapping step, the procedure is not scalable to large designs.
Therefore, we propose an ML-enhanced timing-aware optimization approach to speed up logic optimization by building a predictor for the circuit delay after technology mapping and timing analysis.
This approach is shown to be fast and capable of achieving optimization results of similar quality as the ground truth based on post-mapping delay. Our work makes the following contributions:
\begin{enumerate}
\item 
We show that the correlation between the proxy of maximum delay and post-mapping maximum delay is poor.
\item
We demonstrate that a logic optimization flow driven by post-mapping delay leads to better outcomes than using proxy metrics, but the runtime can be 20$\times$ larger.
\item
We propose an ML-enhanced optimization flow that significantly speeds up the logic optimization driven by post-mapping delay, while maintaining the quality of solution, by incorporating ML inference in the cost calculation.
\end{enumerate}


%% file: sec/2-preliminary.tex
\section{Background}
\label{sec:background}

\begin{figure}[t]
\centering
\includegraphics[width=0.6\linewidth]{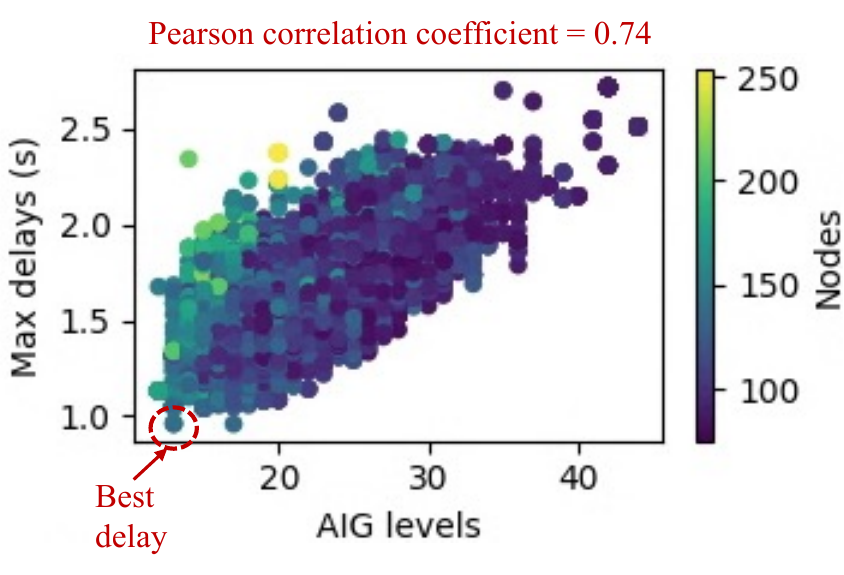}
\caption{Scatter plot: post-mapping circuit delay vs. the number of AIG levels.}
\label{fig:delay-level}
\end{figure}

\noindent
Logic synthesis is a process that turns an abstract specification of desired circuit behavior into a design implementation in terms of logic gates. It involves logic optimization, technology mapping, and post-mapping optimization. The AIG, which decomposes the netlist into an elemental network of AND and Inverter functions, is one of most widely-used initial representations of a netlist, and facilitates the application of structural optimizations to the circuit. The goal of AIG optimization is to apply a sequence of transformations to optimize the power performance and area (PPA) of the circuit, while maintaining the equivalence of the logic to the original specification. Intermediate AIG optimization steps often rely on proxy metrics to estimate final timing and area of the circuit: specifically, the node count of the AIG is used to approximate the design area, and the number of logic levels in the AIG is used to estimate maximum delay. However, the PPA after logic synthesis, determined by technology mapping and timing analysis on gate-level netlist, 
can be inconsistent with the PPA predicted by these proxy metrics. Thus, the design may not be effectively optimized based on inaccurate PPA estimates. 

\input{sec/tbl-aig-structure}

\subsection{Limitations of Optimizing Proxy Metrics}

\noindent
As mentioned above, the correlation between number of AIG levels and the corresponding maximum delay after mapping is imperfect. Fig.~\ref{fig:delay-level} plots the post-technology-mapping maximum delay against the number of levels, for a number of AIGs associated with a multiplier design. From this experiment, we find that the correlation between maximum delay and the number of AIG levels is only 0.74. Moreover, as seen in the figure, the best delay after mapping does not correspond to an AIG with the smallest number of levels. Moreover, another design with fewer AIG levels that the optimal design has $>1.5\times$ larger delay. These observations indicate that proxy metrics may not accurately guide the AIG optimization to achieve the ``true'' optimal design with the best performance.

These inaccuracies expose another major limitation in the use of AIG-optimization-based proxy metrics: two AIGs with same node count and level may have significantly different delay and area at the post-mapping stage. Table~\ref{tbl:aig-structure-diff} shows these metrics for two representations, AIG1 and AIG2, for the same circuit. An optimizer that uses the number of levels as a proxy for delay would treat them as the same and randomly choose one to continue the optimization. This may cause the search to miss a good candidate and thus lose the opportunity to fully explore the design space.

\subsection{Performance-driven Logic Optimization}

\noindent
To resolve the aforementioned issues, we study a performance-driven logic optimization flow. Instead of relying on proxy metrics, the exact post-mapping delay and area are used directly in the cost function to guide AIG optimization, aiming to achieve a more precise and effective optimization process. To demonstrate the effectiveness of introducing the post-mapping delay and area, we compare the outcomes of two different optimization flows: the \textit{baseline flow}, which drives logic optimization using proxy metrics in the cost function, and a \textit{ground-truth-based flow}, which uses the post-timing delay and area instead of proxy metrics. For a multiplier design, we conducted a hyperparameter sweep for these two logic optimization flows to optimize the AIG. We find that for two AIGs with the same area but from the Pareto-optimal fronts of the two flows, the delay of the AIGs optimized using the ground-truth-based flow can be up to 22.7\% better than those optimized using the baseline flow. This suggests that using exact post-mapping delay and area in the optimization process can lead to a more comprehensive exploration of the design space and result in better delay and area.

\begin{figure}[t]
\centering
\includegraphics[width=0.9\linewidth]{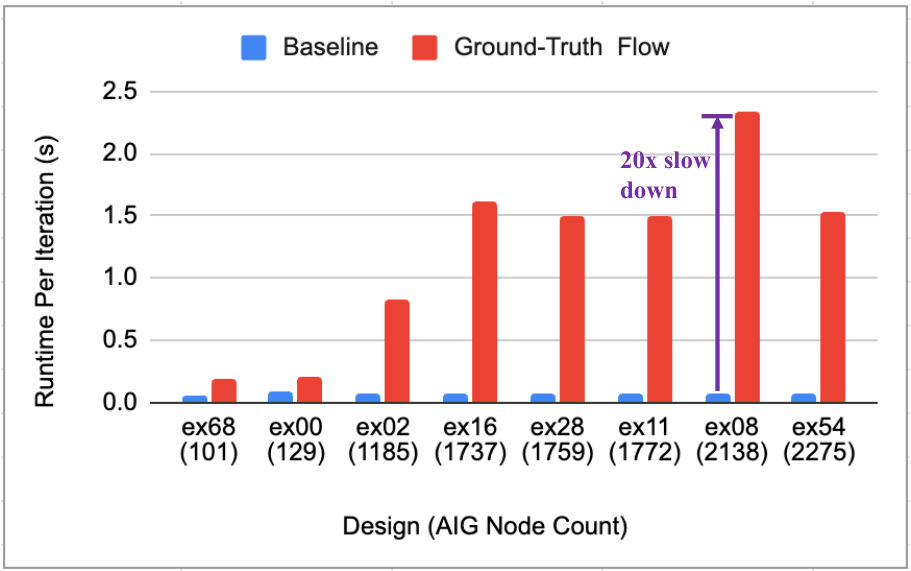}
\caption{Runtime comparison for one iteration in the original logic optimization flow and the ground-truth-based logic optimization flow. The x-axis shows the name of each designs, with the number of its AIG nodes in parentheses.
}
\label{fig:runtime-increase}
\end{figure}

However, ground-truth-based logic optimization requires technology mapping and STA runs during each iteration of the logic optimization flow, which is computationally intensive. Such a flow is found to be up to 20$\times$ slower, on eight designs from IWLS benchmarks~\cite{iwls24}, as illustrated in Fig.~\ref{fig:runtime-increase}.  While the runtimes per iteration are relatively modest, it should be noted that emerging high-quality synthesis approaches, e.g.,~\cite{googleSA23} may require tens and thousands of iterations.
For small-scale circuits, such as ex00 and ex68, even for this large number of iterations the runtime for this ground-truth-based flow may be affordable, and the original flow could be replaced by a ground-truth-based flow to achieve much better PPA. However, for larger designs, the runtime for ground-truth-based optimization becomes prohibitively long.

Thus, it is essential to find an alternative to computing the ground truth delay by running the mapping and static timing analysis (STA) steps.  We propose to build an ML model to predict post-mapping delay, thereby avoiding the need to run mapping and STA in every iteration. This is used to develop an ML-enhanced logic optimization flow in this work, with the aims to achieve AIGs with
comparable quality metrics as ground-truth-based optimization, but with a substantial reduction in runtime.

%% file: sec/tbl-aig-structure.tex
\begin{table}
\centering
\caption{Post-mapping performance for two AIGs with the same number of levels and nodes}
\label{tbl:aig-structure-diff}
\resizebox{0.99\columnwidth}{!}{
\begin{tabular}{|c|c|c|c|c|} 
\hline
\begin{tabular}[c]{@{}c@{}}\textbf{AIG}\\\textbf{Candidate}\end{tabular} & \textbf{Level}      & \begin{tabular}[c]{@{}c@{}}\textbf{Node}\\\textbf{Count}\end{tabular} & \begin{tabular}[c]{@{}c@{}}\textbf{Post-mapping}\\\textbf{Delay (ns)}\end{tabular} & \begin{tabular}[c]{@{}c@{}}\textbf{Post-mapping}\\\textbf{Area ($um^2$)}\end{tabular}  \\ 
\hline
AIG1                                                                     & \multirow{2}{*}{14} & \multirow{2}{*}{178}                                                  & 1.75                                                                               & 803.27                                                                              \\ 
\cline{1-1}\cline{4-5}
AIG2                                                                     &                     &                                                                       & 1.33                                                                               & 770.74                                                                              \\
\hline
\end{tabular}
}
\end{table}

%% file: sec/3-methodology.tex
\section{Methodology}

\subsection{ML-Enhanced Logic Optimization}
\label{sec:ML-enhanced-opt}

\begin{figure}[t]
\centering
\includegraphics[width=\linewidth]{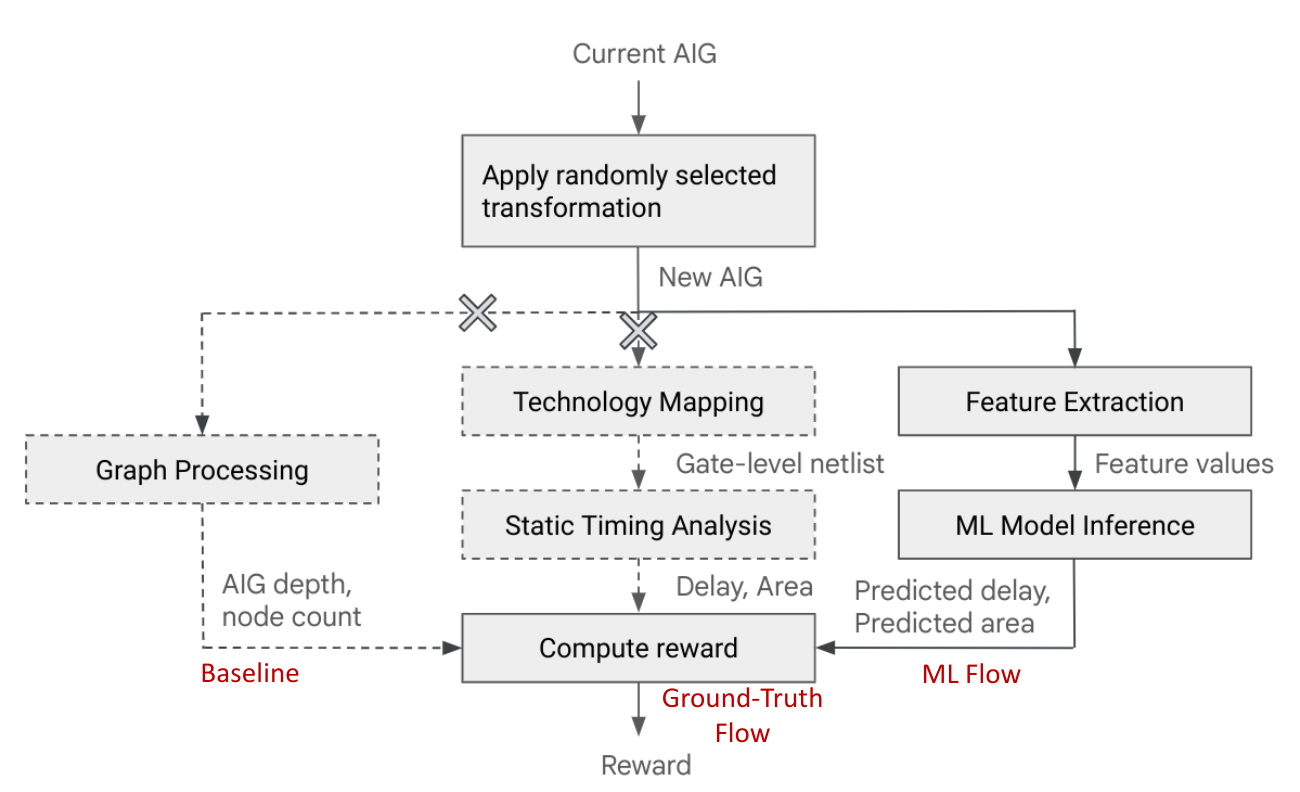}
\caption{Three flows for AIG optimization.}
\label{fig:aig-opt-flow}
\end{figure}

\noindent
Three logic optimization flows discussed in Sec.\ref{sec:background} are illustrated in Fig.~\ref{fig:aig-opt-flow}. From left to right in the figure:


\noindent
(1)~The \textbf{baseline flow} is the original flow with a graph processing step for AIG optimization in which a randomly selected transformation is applied to the AIG at each iteration to produce a new AIG.
An industry flow that we are familiar with uses 103 combinations of the basic transformations~\cite{abc24} available in ABC~\cite{Brayton10}, from which one combination is selected in each iteration and applied to the AIG. The depth and the node count of the new AIG are obtained by processing the graph to evaluate the delay and area in the reward calculation, which determines whether the new AIG will be accepted as the start point for the exploration in the next iteration.

\noindent
(2)~The \textbf{ground-truth-based flow} performs the full technology mapping step, followed by STA, for each new AIG to obtain the post-mapping delay of the mapped gate-level netlist. These precise metrics are used to compute the cost and guide the optimization process. While this method provides accurate PPA results, it is computationally expensive due to the high cost of repeated mapping and STA, and is impractical for large designs.

\noindent
(3)~Our \textbf{proposed ML-based flow} leverages ML models to predict post-mapping delay using features extracted from an AIG. Instead of running mapping and STA in each iteration, a pretrained ML model estimates the delay to guide the optimization, which aims to speed up the flow with high prediction accuracy. A large design, where the ground-truth-based flow incurs large runtimes, can greatly benefit from this flow.

\subsection{Feature Engineering}

\begin{figure}[t]
\centering
\includegraphics[width=0.8\linewidth]{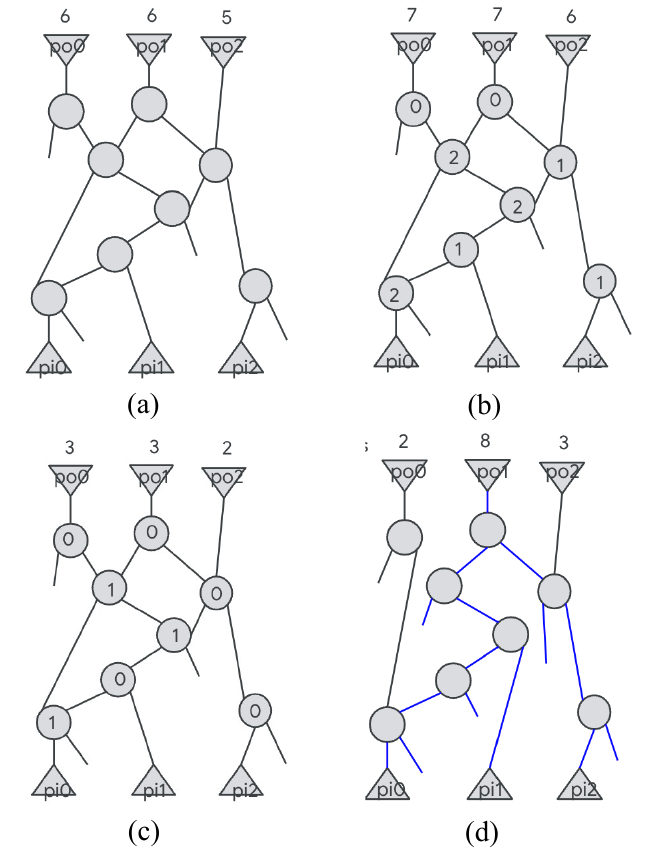}
\caption{Feature extraction in an example AIG. (a) The maximum depth is annotated for each PO. (b) The fanout of each node is annotated as the weight and the updated maximum depth is annotated for each PO. (c) The binary-encoded fanout of each node is annotated as the weight and the updated maximum depth is annotated for each PO. (d) The subgraph of PO1 is highlighed in blue, and the number of paths is annotated for each PO. 
}
\label{fig:feature-demo}
\end{figure}

\noindent
We now enumerate the features used in our approach to drive a decision tree ML model. In principle, it is possible to use graph neural networks (GNNs), which are good at analyzing complex data. However, as the number of features for each node in AIG is limited, GNNs 
are unable to outperform decision-tree-based models on predicting the maximum delay of an AIG. Our experiments find that not only is the GNN-based timing prediction is 2\% worse than decision-tree-based model on average across the designs used in our experiment, but the training cost is also much higher than the lightweight decision-tree-based model. In the context of AIGs, the features available at each node, such as fanin, fanout, or logic type, are relatively simple, and do not fully leverage the strengths of GNNs, and this explains their inability to deliver better performance than decision tree models. Moreover, the maximum delay of a graph are often dominated by several long path and is not greatly affected by the remaining paths, which is hard for a GNN to learn based on its paradigm of message passing.
Thus, we choose decision-tree-based model and extract graph-level features, which is more efficient in learning than relying on node-to-node interactions in our case, for model training. 

The features extracted for each AIG are summarized in Table~\ref{tbl:feature-extraction}, and are based on an analysis of the sources of timing miscorrelation between AIG depth and the maximum delay of the mapped netlist. Two primary sources of this miscorrelation are considered in this work: (a)~the path depth change in the mapped netlist compared to the AIG structure, which impacts the number of stages in the critical path, thus affecting the overall path delay; and (b)~the fanout change after mapping, in which several nodes in AIG are merged into a large cell, resulting in changes to the fanout of the node driving these cells, which affects the gate delay and contributes to the miscorrelation.
The features extracted from AIG in this work can be classified into three categories: the features associated with the critical path, the features related to the fanout distribution, and the features related to the structural complexity of the subgraph correponding to each primary output (PO) reflecting how the topology may affect timing after mapping. Fig.~\ref{fig:feature-demo} demonstrates how the features are extracted using an example AIG with paths between three primary inputs (PIs) and three POs. Different types of depth calculation for PO are shown in Fig.~\ref{fig:feature-demo}(a)-(c), which take the nodes between PO and PI, including the node for PI and excluding the PO node (which is a gate output). Fig.~\ref{fig:feature-demo}(d) shows an example of subgraph extraction for PO.

The features associated with the critical path include three types of features related to AIG depth.

\noindent
\textbf{aig\_nth\_long\_path\_depth}: For each PO, the largest depth is obtained by traversing the graph. Fig.~\ref{fig:feature-demo}(a) annotates each PO by listing the maximum depth from a PI.
In our experiments, we take the top $n$ maximum depths among all POs as our features. The value for parameter $n$ is included in the ``Comments'' column of  Table~\ref{tbl:feature-extraction}.

\noindent
\textbf{aig\_nth\_weighted\_path\_depth}: The feature takes the impact of fanout into account on potential critical paths. The fanout of each node is annotated as the weight of the node, and the graph is traversed to compute the largest depth for each PO with these weights, i.e., with the fanout considered. The top $n$
weighted depths are then used as features.

\noindent
\textbf{aig\_nth\_binary\_weighted\_path\_depth}: The feature considers the probability of nodes being merged into large cells during mapping, which can reduce the length of some paths and reduce delay consequently. Nodes with high fanout have a lower probability of being merged into a large cell. In our approach, nodes with two or more fanouts are assigned a weight of 1, while nodes with fewer than two fanouts are given a weight of 0. Then the largest depth for each PO is updated. The binary weights help model the potential reduction in path depth after mapping.
In our experiments, we empicically set the parameter $n$ in this path-depth-related feature to 3.

\noindent
\textbf{fanout\_mean, max, std, sum}: For the features related to fanout, high-fanout nodes are likely to have high load capacitance, which results in large gate delay. Therefore, we include these features related to the fanout distribution over the graph. These features quantify the overall fanout distribution throughout the AIG by computing the mean, maximum, standard deviation, and sum of fanouts for all nodes. If fanout is uneven across the netlist, with certain regions having much higher fanout than others, the paths with higher fanout are more likely to dominate the overall delay.

\noindent
\textbf{long\_path\_fanout\_mean, max, std, sum}: In a similar way, we gain insights into how fanout distributes at different stages over a long path. If fanout is unevenly distributed along the path, the large delays at the high-fanout stages will result in large path delay due to their higher load capacitance.

To capture the complexity of the paths to each PO, the number of paths is considered by traversing the subgraph of each PO.

\noindent
\textbf{number\_of\_paths}: This feature approximates the probability of a PO having multiple critical and near-critical paths and avoids explicitly enumerating all near-critical paths which is a computationally expensive step for feature extraction.

\input{sec/tbl-features}

\subsection{Data Generation and Model Training}

\noindent
We collect data from eight IWLS benchmarks~\cite{iwls24} for both training and testing. For each design, we generate 40,000 unique AIGs by randomly applying a series of logic transformations available in ABC~\cite{Brayton10}, a widely adopted open-source logic synthesis framework, to build an initial AIG representation of the design. The maximum delay labels are generated by performing technology mapping and STA under a 130nm technology~\cite{skywater130}, using ABC to map each AIG to a standard cell library.

The ML models are implemented using XGBoost~\cite{Chen16}, an ensemble learning algorithm based on gradient boosting. The model is trained using root mean squared error (RMSE) as the loss function. The hyperparameter values are chosen based on grid search. For the XGBoost regressor, we use a learning rate 0.01. We choose the maximum tree depth = 16, the number of estimators = 5000, and the subsampling ratio = 0.8.

%% file: sec/tbl-features.tex
\begin{table*}
\centering
\caption{Features extracted from the AIG}
\label{tbl:feature-extraction}
\resizebox{0.7\textwidth}{!}{
\begin{tabular}{|c|c|} 
\hhline{|==|}
\textbf{AIG features}                   & \textbf{Comments}                                                                                                                                                        \\ 
\hhline{|==|}
number\_of\_node                        & Number of nodes in AIG                                                                                                                                                   \\ 
\hline
aig\_level                              & Level of AIG graph                                                                                                                                                       \\ 
\hhline{|==|}
aig\_nth\_long\_path\_depth             & \begin{tabular}[c]{@{}c@{}}The $n^{\rm th}$ max depth of all POs\\($n=1,2,3$ in experiments)\end{tabular}                                                                          \\ 
\hline
aig\_nth\_weighted\_path\_depth         & \begin{tabular}[c]{@{}c@{}}The $n^{\rm th}$ max depth weighted \\by node fanout of all POs\\($n=1,2,3$ in experiments)\end{tabular}                                                \\ 
\hline
aig\_nth\_binary\_weighted\_path\_depth & \begin{tabular}[c]{@{}c@{}}The $n^{\rm th}$ max depth weighted by~\\0 (when the node inputs 2),\\1 (when the node inputs = 2) of all POs\\($n=1,2,3$ in experiments)\end{tabular}  \\ 
\hhline{|==|}
fanout\_mean, max, std, sum             & \begin{tabular}[c]{@{}c@{}}Mean, max, standard deviation, \\and sum of the fanout of all nodes\end{tabular}                                                              \\ 
\hline
long\_path\_fanout\_mean, max, std, sum & \begin{tabular}[c]{@{}c@{}}Mean, max, standard deviation, \\and sum of fanout of nodes \\on long path (path\_depth = aig\_level)\end{tabular}                            \\ 
\hhline{|==|}
num\_of\_paths                          & Number of paths of each PO (choose top n largest)~                                                                                                                               \\
\hline
\end{tabular}
}
\end{table*}

%% file: sec/4-experiment.tex
\section{Experimental Setup and Results}

\noindent
Our experiments are conducted on eight designs from the IWLS benchmark suite, each from a different functional category and with more than three POs, minimizing similarity between designs and ensuring diversity of the data. Designs with fewer than three POs tend to be simpler and can be handled efficiently without using ML inference, and are therefore not considered in our experiments. The first three columns in Table~\ref{tbl:model-accuracy} summarize the design names, the number of PIs and POs, the median number of AIG nodes across 40K generated AIGs for each design, which range from 69 to 2290 nodes; the precise range for the set of AIGs for each benchmark is shown in the third column of the table.
Four designs are used for model training, and four designs are used for testing to evaluate the ability of the mode to be generalized to unseen designs. We evaluate the prediction accuracy of the model across all designs and compare the ML-based SA logic optimization flow against both the baseline flow and the ground-truth-based flow. The experiments are conducted based on simulated annealing (SA) paradigm which has been applied for circuit optimization with ML in \cite{googleSA23}. Our models can also be integrated into other conventional approaches besides SA. In this work, we choose SA considering two main factors: 1) compared to deterministic algorithms, SA allows to accept temporary cost-increasing solutions with a certain probability during the search process, allowing ``hill-climbing'' that can enable the optimization to potentially find better solutions later; 2) The SA implementation allows the designer to customize methods for estimating the PPA and for tuning the weights for each components in cost function, making it more versatile in handling complex trade-offs.  All runtimes are reported on an AMD EPYC 7B13 CPU @2.3GHz. 

\subsection{Evaluation of Model Accuracy}
\input{sec/tbl-accuracy}

\noindent
Table~\ref{tbl:model-accuracy} summarizes the metrics for the ML model and its prediction accuracy across all designs. The metrics used for evaluation include the mean, maximum, and standard deviation of the absolute \%error as metrics for evaluation, where the absolute \%error is the absolute difference between the ground-truth and the predicted value with respect to the ground-truth value. The results show that the average prediction error across all designs is 4.03\%, demonstrating good overall accuracy, and the average standard deviation is 3.27\% across the designs, so the prediction for most graphs are within a small error.

\subsection{Evaluation of ML-Enhanced Logic Optimization}

\begin{figure}[t]
\centering
\includegraphics[width=0.7\linewidth]{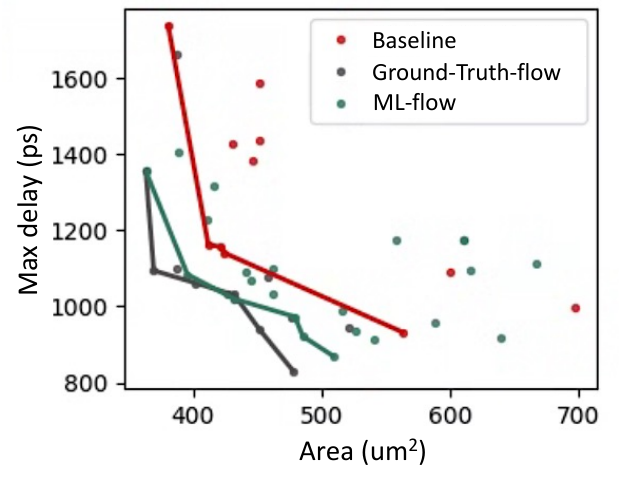}
\caption{A comparison of the Pareto-optimal fronts for the delay and area of a test design from the baseline, ground-truth-based, and ML-based flows. 
}
\label{fig:sol-three-flows}
\end{figure}
\input{sec/tbl-runtime}

\noindent
The three flows described in Section~\ref{sec:ML-enhanced-opt} are applied to the eight IWLS benchmarks and evaluated in terms of both their quality of solution and runtime. Fig.~\ref{fig:sol-three-flows} shows the optimal AIGs from these three flows. The red dots represent the outcomes from baseline optimization flow, the black dots represent the outcomes from the ground-truth-based flow, and the green dots represent the outcomes from the ML-based flow. Each point on the plot corresponds to an optimal AIG obtained from a specific run of SA logic optimization flow with a certain hyperparameter setting. We sweep the hyperparameters to obtain an optimal AIG for different flow settings, which involves sweeping relative weights in cost function and sweep the annealing temperature decay rate. For each flow, a Pareto-optimal curve is generated and shown in the figure. The green curve for the ML-based flow is very closed to the black curve for the ground-truth-based flow, demonstrating that the ML-enhanced approach achieves nearly the same level of quality in terms of delay and area. Both the green and black curves are significantly better than the red curve for the baseline optimization flow using conventional proxy metrics, showing that the ML-based and ground-truth-based flows outperform the original approach in exploring and identifying better designs.

Table~\ref{tbl:sa-runtime} summarizes the runtime of single iteration of the three different flows. The runtime for the baseline flow includes the time elapsed for applying AIG transformation, graph processing to obtain the depth and the node count of the new AIG to evaluate the cost. The runtime for ground-truth-based flow requires a large additional runtime overhead for performing technology mapping and STA on the new AIG. In contrast, the ML-based flow needs a small amount additional runtime for feature extraction and ML inference on the new AIG, but avoids the costly mapping and STA. The ML-based flow achieves a significant runtime reduction of 80.83\% compared to the ground-truth-based flow on average and  maximum 88.79\% reduction across all testcases, while delivering similar quality of solution. This demonstrates that the ML-enhanced logic optimization flow can not only maintain high-quality results but also make substantial improvements on efficiency, which makes it an excellent alternative for larger designs.

%% file: sec/tbl-accuracy.tex
\begin{table}
\centering
\caption{Accuracy of XGBoost model for timing prediction 
}
\label{tbl:model-accuracy}
\resizebox{0.99\columnwidth}{!}{
\begin{tabular}{|c|c|c|c|c|c|c|} 
\hline
\multicolumn{2}{|c|}{\textbf{Design}}           & \textbf{PI/PO} & \begin{tabular}[c]{@{}c@{}}\textbf{\#Node}\\\textbf{(Range)}\end{tabular} & \begin{tabular}[c]{@{}c@{}}\textbf{Mean of }\\\textbf{\% Error}\end{tabular} & \begin{tabular}[c]{@{}c@{}}\textbf{Max of}\\\textbf{\%Error}\end{tabular} & \begin{tabular}[c]{@{}c@{}}\textbf{Std. of}\\\textbf{\% Error}\end{tabular}  \\ 
\hline
\multirow{4}{*}{\textbf{\rotatebox{90}{Training}}} & \textbf{EX00} & 16/7           & 69-189                                                                  & 4.24\%                                                                       & 29.87\%                                                                   & 3.56\%                                                                       \\ 
\cline{2-7}
                                & \textbf{EX08} & 18/5           & 1828-1448                                                               & 1.90\%                                                                       & 14.05\%                                                                   & 1.59\%                                                                       \\ 
\cline{2-7}
                                & \textbf{EX28} & 17/7           & 1296-2222                                                               & 1.53\%                                                                       & 14.41\%                                                                   & 1.36\%                                                                       \\ 
\cline{2-7}
                                & \textbf{EX68} & 14/7           & 62-140                                                                  & 4.23\%                                                                       & 33.64\%                                                                   & 3.48\%                                                                       \\ 
\hline
\multirow{4}{*}{\textbf{\rotatebox{90}{Test}}}  & \textbf{EX02} & 18/6           & 848-1522                                                                & 5.77\%                                                                       & 32.52\%                                                                   & 4.86\%                                                                       \\ 
\cline{2-7}
                                & \textbf{EX11} & 17/7           & 1253-2290                                                               & 5.22\%                                                                       & 36.96\%                                                                   & 3.90\%                                                                       \\ 
\cline{2-7}
                                & \textbf{EX16} & 16/5           & 1237-2236                                                               & 4.50\%                                                                       & 36.74\%                                                                   & 3.55\%                                                                       \\ 
\cline{2-7}
                                & \textbf{EX54} & 17/7           & 1469-3080                                                               & 4.83\%                                                                       & 39.85\%                                                                   & 3.87\%                                                                       \\ 
\hline
~                               & ~             & ~              & ~                                                                       & Avg. 4.03\%                                                                  & Max 39.85\%                                                               & Ave. 3.27\%                                                                  \\
\hline
\end{tabular}
}
\end{table}

%% file: sec/tbl-runtime.tex
\begin{table}
\centering
\caption{Runtime for the three flows 
}
\label{tbl:sa-runtime}
\resizebox{0.99\columnwidth}{!}{
\begin{tabular}{|c|c|c|c|c|} 
\hline
\multicolumn{2}{|c|}{\textbf{Design}}           & \textbf{Baseline (s)} & \begin{tabular}[c]{@{}c@{}}\textbf{Ground-Truth-flow}\\\textbf{Mapping+STA (s)}\end{tabular} & \begin{tabular}[c]{@{}c@{}}\textbf{ML-flow}\\\textbf{ML Inference (s)}\end{tabular}  \\ 
\hline
\multirow{4}{*}{\textbf{\rotatebox{90}{Training}}} & \textbf{EX00} & 0.0936                & 0.3028                                                                             & 0.1118 (-63.08\%)                                                                    \\ 
\cline{2-5}
                                & \textbf{EX08} & 0.0736                & 2.3375                                                                             & 0.1967 (-88.79\%)                                                                    \\ 
\cline{2-5}
                                & \textbf{EX28} & 0.0712                & 1.4966                                                                             & 0.1555 (-85.54\%)                                                                    \\ 
\cline{2-5}
                                & \textbf{EX68} & 0.0463                & 0.1961                                                                             & 0.0166 (-74.05\%)                                                                    \\ 
\hline
\multirow{4}{*}{\textbf{\rotatebox{90}{Test}}}  & \textbf{EX02} & 0.0670                & 0.8335                                                                             & 0.1213 (-79.09\%)                                                                    \\ 
\cline{2-5}
                                & \textbf{EX11} & 0.0693                & 1.5044                                                                             & 0.1575 (-85.59\%)                                                                    \\ 
\cline{2-5}
                                & \textbf{EX16} & 0.0764                & 1.6131                                                                             & 0.1617 (-85.91\%)                                                                    \\ 
\cline{2-5}
                                & \textbf{EX54} & 0.0766                & 1.5320                                                                             & 0.1715 (-84.58\%)                                                                    \\ 
\hline
\multicolumn{2}{|c|}{~ Avg.}                    & ~                     & ~                                                                                  & -80.83\%                                                                             \\ 
\hline
\multicolumn{2}{|c|}{Max}                       &                       &                                                                                    & -88.79\%                                                                             \\
\hline
\end{tabular}
}
\end{table}

%% file: sec/5-conclusion.tex
\section{Conclusion}

\noindent
This paper proposed an ML-enhanced performance-driven logic optimization flow that addresses the limitations of conventional logic optimization flow 
relying on proxy metric for delay estimation. While a ground-truth-based flow, incorporating exact post-mapping delay, improves design quality, it significantly increases runtime. Our experimental results show that our approach offers a practical solution for large designs, which generates design of better quality with small runtime overhead.